\documentclass[12pt, draftcls, onecolumn]{IEEEtran}
\ifCLASSINFOpdf
  \usepackage[pdftex]{graphicx}
  \DeclareGraphicsExtensions{.pdf,.jpeg,.png}
\else
  \usepackage[dvips]{graphicx}
\fi
\usepackage{epstopdf}
\usepackage[cmex10]{amsmath}
\usepackage{amssymb}
\usepackage{wasysym}
\usepackage{algorithm}
\usepackage{algorithmic}
\usepackage{array}
\usepackage{cite}
\usepackage{color}
\usepackage{url}

\usepackage{epsfig,latexsym}
\usepackage{cite}
\usepackage{verbatim}
\usepackage{amsopn}
\usepackage{booktabs}
\usepackage{bm}
\usepackage{stfloats}
\usepackage{hyperref}
\usepackage{graphicx}
\usepackage{float}
\usepackage{subfigure}
\usepackage{makecell}

\begin{document}

\title{Channel Estimation for Wideband MmWave MIMO OFDM System Exploiting Block Sparsity}

\author {Yujie~Wang,~\IEEEmembership{Student~Member,~IEEE}, Chenhao~Qi,~\IEEEmembership{Senior~Member,~IEEE},  \\ Ping~Li, Zhaohua~Lu, and Ping~Lu
\thanks{Yujie~Wang and Chenhao~Qi are with the School of Information Science and Engineering, Southeast University, Nanjing 210096, China (Email: qch@seu.edu.cn).}
\thanks{Ping~Li, Zhaohua~Lu, and Ping~Lu are with ZTE Corporation, Shenzhen 518057, China.}
}

\markboth{}
{}

\maketitle

\begin{abstract}
In this letter, we investigate time-domain channel estimation for wideband millimeter wave (mmWave) MIMO OFDM system. By transmitting frequency-domain pilot symbols as well as different beamforming vectors, we observe that the time-domain mmWave MIMO channels exhibit channel delay sparsity and especially block sparsity among different spatial directions. Then we propose a time-domain channel estimation exploiting block sparsity (TDCEBS) scheme, which always aims at finding the best nonzero block achieving the largest projection of the residue at each iterations. In particular, we evaluate the system performance using the QuaDRiGa which is recommended by 5G New Radio to generate wideband mmWave MIMO channels. The effectiveness of the proposed TDCEBS scheme is verified by the simulation results, as the proposed scheme outperforms the existing schemes. 
\end{abstract}

\begin{IEEEkeywords}
5G, channel estimation, mmWave communications, OFDM, sparse recovery.
\end{IEEEkeywords}

\section{Introduction}
Since the fifth generation (5G) wireless communications has been standardized in 2020s, more concern is gathered by the industry, academia and government on beyond 5G (B5G) or even 6G. To support high data rate transmission, wireless communications working on millimeter wave (mmWave) frequency band is on focus. Early works on mmWave MIMO communications and signal processing consider the narrow-band mmWave channels, by assuming the delay of different channel paths is the same~\cite{heath2016overview, yang2019hardware, ma2020sparse}. But later on, practical testing results show that the mmWave channels are wideband, where the delay of different channel paths are different~\cite{wideband,kim2021spatial,Gao2016HDW}. Since the delay spread causes frequency-selective fading, orthogonal frequency division multiplexing (OFDM) is introduced to mmWave MIMO communications, just like in sub-6GHz MIMO communications~\cite{shu2017pilot}.

To acquire accurate channel state information for mmWave MIMO beamforming, efficient channel estimation is needed. The main-stream methods for wideband mmWave MIMO can be divided into two categories, including frequency-domain channel estimation methods~\cite{heath2018freq,you2020tensor} and time-domain channel estimation methods~\cite{kim2019twostep,gred2019SBL}. In~\cite{heath2018freq}, a simultaneous weighted orthogonal matching pursuit (SW-OMP) channel estimation method is proposed to estimate the frequency-domain mmWave MIMO channels by exploiting the spatial sparsity that is also called angle sparsity sometimes. In~\cite{you2020tensor}, the received training signal is treated as a low-rank three-dimensional tensor that fits a canonical polyadic model, where a structured canonical polyadic decomposition-based channel estimation method is proposed by utilizing the Vandermonde property of factor matrices that contains the channel parameters. Compared to frequency-domain channel estimation, time-domain channel estimation is challenging, since the standard least squared (LS) methods cannot be applied due to the low rank issue. But on the other hand, time-domain channel estimation can further exploit the sparse property of the channel delay spread. In~\cite{kim2019twostep}, a two-step time-domain channel estimation scheme is proposed, where the effective channel composed of beamformer, channel steering vectors, and pulse shaping is firstly estimated by the LS method, and then the desired channel is estimated by the orthogonal matching pursuit (OMP) algorithm. But the block sparsity is not considered by~\cite{kim2019twostep}. In~\cite{gred2019SBL}, a block sparse channel estimation method based on sparse Bayesian learning (SBL) is proposed by exploiting the channel delay sparsity and block sparsity among different angles.

In this letter, we consider the time-domain channel estimation for wideband mmWave MIMO OFDM system. By transmitting frequency-domain pilot symbols as well as different beamforming vectors, we observe that the time-domain mmWave MIMO channels exhibit channel delay sparsity and especially block sparsity among different spatial directions. Then we propose a time-domain channel estimation exploiting block sparsity (TDCEBS) scheme, which always aims at finding the best nonzero block achieving the largest projection of the residue at each iterations. In particular, we evaluate the system performance using the QuaDRiGa which is recommended by 5G New Radio to generate wideband mmWave channels.

The notations used in this letter are as follows. Symbols for matrices and vectors are written in boldface. $a, \bm{a}, \bm{A}$ denote a scalar, a vector and a matrix, respectively, while $(*)^T, (*)^H, \|*\|_2$ denote the transpose, the conjugate transpose and the $\ell_2$ norm, respectively. $\bm A(m,n)$ represents the entry located in the $m$th row and $n$th column of matrix $\bm A$. $\bm{I}_K$ denotes a $K$-by-$K$ identity matrix. $\mathcal{CN}(m, \bm{R})$ represents the complex Gaussian distribution whose mean is $m$ and covariance matrix is $\bm{R}$. $\mathbb{C}$ and $\emptyset$ represent the set of complex-valued numbers and the empty set, respectively. $\cup$ denotes the union of sets.

\section{System Model}\label{sec.system.model}
\begin{figure}[!t]
	\centerline{\includegraphics[height=4.6 cm]{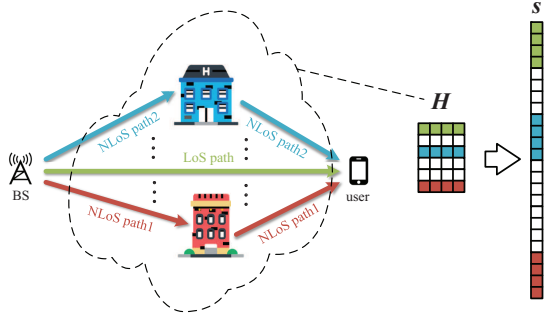}}
	\caption{Wideband mmWave MIMO system with block-sparse channels.}
	\label{channelStructure}
\end{figure}

We consider a downlink mmWave MIMO OFDM system, as shown in Fig.~\ref{channelStructure}. The base station (BS) is equipped with $N_ {\rm{BS}}$ antennas in uniform linear arrays. To simplify the analysis, we assume that the users are equipped with a single antenna. To tackle the frequency-selective fading in wideband mmWave channels, OFDM is typically used for the signal transmission. The total number of OFDM subcarriers is denoted by $N_{\rm c}$ and the subcarriers spacing is denoted by $\Delta f$. Then the duration of each OFDM symbol is $ 1/\Delta f$, which is further divided into $N_{\rm c}$ OFDM samples and the duration of each OFDM sample is
\begin{equation}\label{sampleDuration}
  T_{\rm s}=\frac{1}{N_{\rm c} \Delta f}.
\end{equation}
In order to eliminate inter-symbol interference and inter-carrier interference, a cyclic prefix (CP) is placed at the head of each OFDM symbol. We denote the length of the CP by $N_{\rm cp}$.

We denote the number of resolvable channel paths for the mmWave MIMO system by $L$, including a line-of-sight (LoS) path and $L-1$ non-line-of-sight (NLoS) paths, as shown in Fig.~\ref{channelStructure}. Generally, the LoS path is the path with the largest transmission gain. However, in some scenarios such as NLoS scenario, since the LoS path is blocked by channel obstacles, the transmission gain of the LoS path may be reduced to be comparable to that of the NLoS path, or even smaller than that of the NLoS path. That is to say, the LoS path cannot provide effective transmission gain and therefore can be treated a NLoS path. 
We denote the channel gain, delay and angle-of-departure (AoD) of each channel path respectively by $\gamma_l$, ${\tau_l}$, and $\theta_l$, for $l=1,2,\ldots,{L}$. With these channel parameters, the Saleh-Valenzuela channel model widely used in the narrowband mmWave MIMO systems can be extended to the wideband mmWave MIMO systems~\cite{ma2020high}. Accordingly, the wideband mmWave MIMO channel can be written as 
\begin{equation}\label{widebandChannel}
  \bm {h}(t) = \sqrt{\frac{N_ {\rm{BS}}}{L}}\sum_{l = 1}^{L}\gamma_l p(t-\tau_l) \bm {\alpha}^H(N_{\rm BS}, \theta_l),
\end{equation}
where $p(t)$ denotes the pulse shaping function and $\bm \alpha(N_{\rm BS}, \theta_l)$ denotes the channel steering vector expressed as
\begin{equation}\label{ChannelSteeringVector1}
\bm{\alpha}(N_{\rm BS}, {\theta_l})=\frac{1}{\sqrt{N_{\rm BS}}}[1,e^{j\pi\theta_l},\ldots,e^{j(N_{\rm BS}-1)\pi\theta_l}]^T.
\end{equation}
After channel sampling, \eqref{widebandChannel} can be written as
\begin{equation}\label{sampledChannel}
  \bm {h}(n) = \sqrt{\frac{N_ {\rm{BS}}}{L}}\sum_{l = 1}^{L}\gamma_l p(nT_{\rm s}-\tau_l) \bm {\alpha}^H(N_{\rm BS}, \theta_l),
\end{equation}
for $n = 0,1\ldots,N_{\rm{cp}}-1$, since the OFDM modulation requires the CP length larger than the maximum channel delay spread. We define a time-domain channel matrix $\bm{H}\in\mathbb{C}^{N_{\rm cp}\times N_{\rm BS}}$ as
\begin{equation}
	\bm{H}\triangleq \big\{\bm{h}(0)^T,\bm{h}(1)^T,\ldots,\bm{h}(N_{\rm cp}-1)^T \big\}^T.
\end{equation}

To estimate $\bm{H}$, we transmit frequency-domain pilot symbols with the length of $K (K\leq N_{\rm c})$, where the $k$th transmitted pilot symbol is denoted by $x(k)$ and the corresponding received pilot symbol is denoted by $y(k),~k=1,2,\ldots,K$. Then we define the received pilot vector as
\begin{equation}
	\bm{y}\triangleq [y(1),y(2),\ldots,y(K)]^T.
\end{equation}
We have
\begin{equation}\label{pilotSignal}
  \bm{y} = \bm{X} \bm{D} \bm{H} \bm{f} + \bm{\eta},
\end{equation}
where $\bm {X} \in \mathbb{C}^{K \times K}$ is a diagonal matrix whose $k$th diagonal entry is $x(k)$, and $\bm {\eta} \in\mathbb{C}^{K}$ represents an additive white Gaussian noise (AWGN) vector with
$\bm{\eta}\sim\mathcal{CN}({\bm 0},\sigma^2\bm{I}_K)$. $\bm {D} \in \mathbb{C}^{ K \times N_{\rm cp}}$ is a submatrix of the standard discrete Fourier transform (DFT) matrix. Given the standard DFT matrix $\bm{G}\in \mathbb{C}^{ N_{\rm c} \times N_{\rm c}}$, we determine $\bm{D}$ by extracting the first $N_{\rm cp}$ columns from $\bm{G}$ and selecting the $K$ rows corresponding to the pilot subcarriers from $\bm{G}$. In~\eqref{pilotSignal}, $\bm {f}\in \mathbb{C}^{N_{\rm BS}}$ is a beamforming vector to form directional signal transmission from the BS to the user.

To scan the whole angle space that the signal of the BS covers, we use $N_{\rm b}$ different beamforming vectors for pilot transmission. Note that if we only use one beam, it may occur that the beam is not aligned with the effective channel path and results in weak received signal strength and consequently poor channel estimation performance~\cite{qi2020codebook}. The $N_{\rm b}$ different beamforming vectors form a codebook matrix
\begin{equation}
\bm {F} \triangleq [\bm {f}_1, \bm {f}_2,\ldots,\bm {f}_{N_{\rm b}}] \in \mathbb{C}^{N_{\rm BS} \times N_{\rm b}}.
\end{equation}
Typically we need $N_{\rm b} \geq N_{\rm BS}$. Based on \eqref{pilotSignal}, we have
\begin{equation}\label{signalMatrix}
  \bm {Y} = \bm{X} \bm{D} \bm{H} \bm{F} + \bm{\Psi},
\end{equation}
where $\bm{Y}\triangleq [\bm{y}_1,\bm{y}_2,\ldots,\bm{y}_{N_{\rm b}}] \in \mathbb{C}^{K \times N_{\rm b}}$ is the received pilot matrix after using $N_{\rm b}$ different beamforming vectors, and $\bm{\Psi}$ is the consequent channel noise matrix.

\section{Channel Estimation Exploiting Block Sparsity}\label{sec.motivation}
By multiplying the right pseudo inverse of $\boldsymbol{F}$, denoted as $\bm {F}^{H}(\bm {F}\bm {F}^H)^{-1} $, on both sides of \eqref{signalMatrix}, we have
\begin{equation}\label{signalMatrixAfter}
  \bm {Y} \bm {F}^{H}(\bm {F}\bm {F}^H)^{-1} = \bm{X} \bm{D} \bm{H} + \bm{\Psi}\bm {F}^{H}(\bm {F}\bm {F}^H)^{-1}.
\end{equation}
To simplify the notation, we define
\begin{align}
	\bm{A}\triangleq & \bm{X} \bm{D} \in \mathbb{C}^{K \times N_{\rm cp}}, \\
	\bm{Z}\triangleq &\bm{Y} \bm {F}^{H}(\bm {F}\bm {F}^H)^{-1} \in \mathbb{C}^{K \times N_{\rm BS}}, \\
	\bm{N}\triangleq & \bm{\Psi} \bm {F}^{H}(\bm {F}\bm {F}^H)^{-1} \in \mathbb{C}^{K \times N_{\rm BS}}.
\end{align}
Then \eqref{signalMatrixAfter} can be rewritten as
\begin{equation}\label{signalMatrixSparse}
	\bm {Z} = \bm{A} \bm{H} + \bm{N}.
\end{equation}
The LS estimation of $\bm{H}$ can be expressed as
\begin{equation}\label{LSEstimationOfChannel}
  \widehat{\bm{H}} = (\bm{A}^H \bm{A})^{-1}\bm{A}^H \bm {Z}.
\end{equation}
However, \eqref{LSEstimationOfChannel} is based on the premise that $\bm{A}$ is full column rank. In practice, even if $K \geq N_{\rm cp}$, we cannot guarantee that $\bm{A}$ is always full column rank, since the rank of $\bm{A}$ is essentially determined by that of $\bm{D}$. If we use a small number of pilot symbols, i.e., $K$ is small, it will occur with high probability that $\bm{A}$ is a low-rank matrix. If we use $K < N_{\rm cp}$ to reduce the number of pilot symbols and increase the spectral efficiency, $\bm{A}$ will be a low-rank matrix. In this context, \eqref{LSEstimationOfChannel} cannot be used for channel estimation.

\begin{figure}[!t]
	\centerline{\includegraphics[height=6.5 cm]{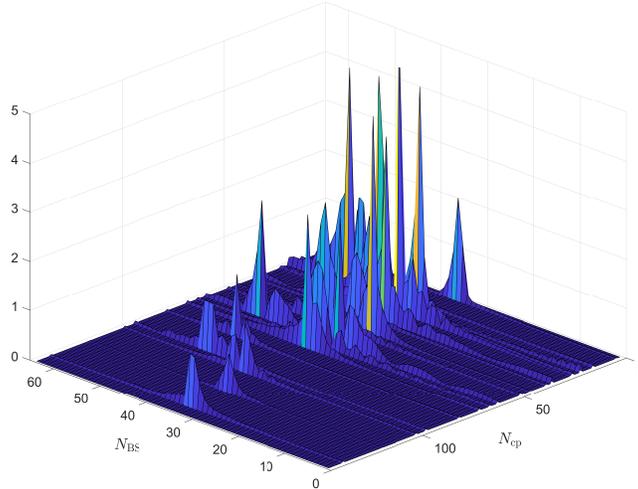}}
	\caption{Time-domain mmWave MIMO channels generated by QuaDRiGa.}
	\label{timeDomainChannel}
\end{figure}

Some recent work on channel modeling and channel measurements shows that mmWave channels are typically sparse~\cite{gred2019SBL, liao2019closed}. In practice, the number of channel paths is much smaller than the length of the maximum channel delay spread, i.e., $L \ll  N_{\rm cp}$, which causes most $\bm{h}(n)$ to be zero vectors for $n=0,1,\ldots, N_{\rm cp}-1$. Most rows of $\bm{H}$ are zero and the number of nonzero rows of $\bm{H}$ is $L$. In Fig.~\ref{timeDomainChannel}, we use the QuaDRiGa which is recommended by 5G New Radio (NR), to generate the time-domain mmWave MIMO channels~\cite{quadriga}, where the channels with $L=21$ paths and maximum delay spread of $N_{\rm cp}=144$ taps are scanned by $N_{\rm b}=N_{\rm BS}=64$ DFT codewords. Note that here we consider the NLoS scenario where the LoS path is blocked and all the 21 paths are treated as NLoS paths. From Fig.~\ref{timeDomainChannel}, it is seen that along each spatial direction that scanned by a DFT codeword, the time-domain channel is a sparse vector with the length of $N_{\rm cp}=144$ taps but only $L=21$ nonzero taps. In particular, $N_{\rm b}=64$ time-domain channel vectors pointing at different spatial directions share the common sparsity, although their channel gain might be substantially different.

Define
\begin{align}
	N_{\rm s}\triangleq & N_{\rm cp} N_{\rm BS}, \\
	N_{\rm q}\triangleq & K N_{\rm BS}.
\end{align}
By string together each row of $\bm{H}$, we convert $\bm{H}$ to be a column vector $\bm{s}$ as
\begin{equation}
	\bm {s} \triangleq \big[\bm {h}(0), \bm {h}(1),\ldots, \bm {h}(N_{\rm cp}-1) \big]^T \in \mathbb{C}^{N_{\rm s}}.
\end{equation}
As illustrated in Fig.~\ref{channelStructure}, $\bm {s}$ exhibits block sparsity, where the length of each block is $N_{\rm BS}$.

Define the $k$th row of $\bm{Z}$ and $\bm{N}$ as $\boldsymbol{z}_k$ and $\boldsymbol{n}_k$, respectively, for $k=1,2,\ldots,K$. We have
\begin{align}
	\bm{Z}=&[\bm{z}_1^T,\bm{z}_2^T,\ldots,\bm{z}_K^T]^T,\\
	\bm{N}=&[\bm{n}_1^T,\bm{n}_2^T,\ldots,\bm{n}_K^T]^T.
\end{align}
Then we define
\begin{align}
	\bm{q} \triangleq & [\bm z_1, \bm z_2,\ldots,\bm z_{K}]^T \in \mathbb{C}^{N_{\rm q}},\\
	\bm{v} \triangleq & [\bm n_1, \bm n_2,\ldots,\bm n_{K}]^T \in \mathbb{C}^{N_{\rm q}}.
\end{align}
We further define a stacked measurement matrix
\begin{equation}\label{stackMeasureMatrix}
	\bm B \triangleq \begin{pmatrix}
		\bm {B}_{1,1} & \bm {B}_{1,2} &\cdots & \bm {B}_{1,N_{\rm cp}} \\
		\bm {B}_{2,1} & \bm {B}_{2,2} & \cdots & \bm {B}_{2,N_{\rm cp}} \\
		\vdots & \vdots & \ddots & \vdots  \\
		\bm {B}_{K,1} & \bm {B}_{K,2} & \cdots & \bm {B}_{K,N_{\rm cp}}
	\end{pmatrix} \in \mathbb{C}^{N_{\rm q}\times N_{\rm s}}
\end{equation}
where the $(k, m)$th block of $\bm B$, denoted as $\bm {B}_{k,m}$, is a diagonal matrix, with the definition of $\bm {B}_{k,m}$ as
\begin{equation}
	\bm {B}_{k,m} \triangleq \bm {A}(k, m) \bm {I}_{N_{\rm BS}} \in \mathbb{C}^{{N_{\rm BS}} \times {N_{\rm BS}} }
\end{equation}
for $k=1,2,\ldots,K$ and $m=1,2,\ldots,N_{\rm cp}$.

Finally, (\ref{signalMatrixSparse}) can be rewritten as
\begin{equation}\label{stackSignalVector}
	\bm {q} = \bm {B}\bm {s} + \bm {v}
\end{equation}

To fully exploit the block sparsity of $\bm{s}$, we can resort to the block sparse recovery algorithms.

First we define a residual vector $\bm r \in \mathbb{C}^{N_{\rm q}}$, which is initialized to be $\bm r= \bm q$. We define a non-negative  integer set $\bm{\Lambda}$ to keep the indices of the selected blocks of $\bm s$ during the iterations, where $\bm{\Lambda}$ is initialized to be an empty set.

At each iteration, we always aim at finding the best nonzero block of $\bm s$, whose corresponding block of $\bm B$ can achieve the largest projection of $\bm r$ among all possible blocks. The projection is usually measured based on the LS estimation. Then the index of the best nonzero block of $\bm s$ can be obtained by
\begin{equation}\label{indexSearch}
  I = \arg\max_{i\in\{ 1,2,\ldots, N_{\rm cp}\}\backslash {\bm \Lambda}} \big\|(\bm {P}_i^H \bm {P}_i)^{-1} \bm {P}^H_i \bm r\big\|_2,
\end{equation}
where ${\bm P}_i \in \mathbb{C}^{N_{\rm q}\times N_{\rm BS}}$ is the $i$th column block of $\bm B$ defined as
\begin{equation}\label{measurementBlock}
  \bm {P}_i \triangleq \big[\bm {B}_{1,i}^T,
  \bm {B}_{2,i}^T,
  \ldots,
  \bm {B}_{K,i}^T\big]^T
\end{equation}
for $i=1,2,\ldots,N_{\rm cp}$.
Since $\bm {B}_{k,m}$ is a diagonal matrix, different columns of $\bm {P}_i$ are mutually orthogonal. In practice, the power of pilot symbols is in general the same. Therefore, $\bm{P}_i^H \bm {P}_i$ is a unit matrix multiplied by a constant. Then (\ref{indexSearch}) can be simplified as
\begin{equation}\label{simpleIndexSearch}
  I = \arg\max_{i\in\{ 1,2,\ldots, N_{\rm cp}\}\backslash{\bm \Lambda}} \big \|\bm {P}^H_i \bm r \big\|_2.
\end{equation}
Once $I$ is selected, it is added to ${\bm \Lambda}$ to update ${\bm \Lambda}$ as
\begin{equation}\label{UpdateSelection}
	{\bm \Lambda} \leftarrow {\bm \Lambda} \cup \{I\}.
\end{equation}
Then we also update the residue vector by removing all projection of $\bm q$ on the selected blocks as
\begin{equation}\label{updateResidual}
  \bm r \leftarrow \bm q - \bm{P}_{\bm \Lambda} (\bm{P}_{\bm \Lambda}^H \bm{P}_{\bm \Lambda})^{-1} \bm{P}_{\bm \Lambda}^H \bm{q}
\end{equation}
where
\begin{equation}
	\bm P_{\bm \Lambda} \triangleq \{ \bm P_i,~i\in{\bm \Lambda} \}
\end{equation}
is a submatrix of $\bm B$ by selecting the column blocks of $\bm B$ indexed by ${\bm \Lambda}$.

We iteratively run \eqref{simpleIndexSearch}, \eqref{UpdateSelection} and \eqref{updateResidual} before satisfying a \textit{stop condition}. Here are three different kinds of setting of the \textit{stop condition}.
\begin{enumerate}
	\item[1)]
We can simply stop the iteration when the number of iterations exceeds the predetermined maximum number of iterations.

\item[2)] We can also set the \textit{stop condition} to be a threshold, since the power of the residue vector is getting smaller with the increasing number of iterations. When $\|\bm r\|_2$ is smaller than the threshold, we stop the iteration.

\item[3)] We may also compare the power of the residue vector between the current iteration and the last iteration. If the power of the residue vector at current iteration is even larger than that of the last iteration, implying that the residue cannot get smaller any longer, we stop the iteration.
\end{enumerate}


\begin{algorithm}[!t]
\caption{Time-domain Channel Estimation Exploiting Block Sparsity (TDCEBS)}
\label{alg1}
    \begin{algorithmic}[1]
        \REQUIRE $\bm {A}$, $\bm Z$.
        \ENSURE $\widehat{\bm {H}}_{\bm \Lambda}$.
        \STATE Initialization: $\bm r \leftarrow \bm q$,  ${\bm \Lambda} \leftarrow \emptyset$.
        \WHILE{\textit {the stop condition is not satisfied}}
        \STATE Obtain the index of the best nonzero block via \eqref{simpleIndexSearch}.
        \STATE Update the indice set via \eqref{UpdateSelection}.
        \STATE Update the residue vector via \eqref{updateResidual}.
        \ENDWHILE
        \STATE Estimate the nonzero row of $\bm{H}$ via \eqref{estimationChannel}.
    \end{algorithmic}
\end{algorithm}

Finally, we estimate the nonzero rows of $\bm H$ based on $\bm \Lambda$ and $\bm Z$ by
\begin{equation}\label{estimationChannel}
  \widehat{\bm {H}}_{\bm \Lambda} = (\bm {A}_{\bm \Lambda}^H \bm {A}_{\bm \Lambda})^{-1}\bm {A}_{\bm \Lambda}^H \bm Z.
\end{equation}
where
\begin{equation}
	\bm A_{\bm \Lambda} \triangleq \{ \bm A_i,~i\in{\bm \Lambda} \}
\end{equation}
is a submatrix of $\bm A$ by selecting the columns of $\bm A$ indexed by ${\bm \Lambda}$. Once the nonzero rows of $\bm H$ is obtained, $\bm H$ is also estimated. The detailed steps of the proposed TDCEBS scheme are summarized in \textbf{Algorithm~\ref{alg1}}.

\section{Simulation Results}\label{sec.simulation}
To evaluate the system performance, we use the QuaDRiGa which is recommended by 5G NR to generate wideband mmWave channels~\cite{quadriga}. We consider a BS equipped with $N_{\rm BS}=64$ antennas working at 28GHz mmWave frequency band serves several single-antenna users. As shown in Fig.~\ref{timeDomainChannel}, the NLoS scenario with $L=21$ channel paths is generated by the QuaDRiGa. For the OFDM modulation in 5G NR high-frequency standard, the IFFT size is $N_{\rm c}=2048$; the CP length is $N_{\rm cp}=144$; and the OFDM subcarrier spacing is $\Delta f= 120 \rm KHz$. We set the \textit{stop condition} to be a threshold 0.01 in the simulation. When $\|\bm r\|_2$ is smaller than 0.01, we stop the iteration. The detailed simulation parameters used for the QuaDRiGa are shown in Table~\ref{parameters}. 
\begin{table}[h]
  \caption{Parameters of QuaDRiGa.}
  \centering
  \begin{tabular}{p{4.5cm}p{3cm}}
    \hline
    \makecell[c]{Parameter} & \makecell[c]{value}\\
    \hline
    \makecell[c]{Number of BS antennas }  & \makecell[c]{${N_{\rm BS}}=64$} \\
    \makecell[c]{Number of paths } & \makecell[c]{$L=21$} \\
    \makecell[c]{Number of subcarriers }& \makecell[c]{$N_{\rm c} = 2048$}\\
    \makecell[c]{CP length }& \makecell[c]{$N_{\rm cp} = 144$}\\
    \makecell[c]{Subcarrier spacing } & \makecell[c]{$\Delta f= 120 \rm KHz$}\\
    \makecell[c]{Center frequency} & \makecell[c]{$28 \rm GHz$} \\
    \makecell[c]{Height of BS} & \makecell[c]{$25 \rm m$} \\
    \makecell[c]{Height of UE} & \makecell[c]{$1.5 \rm m$} \\
    \makecell[c]{Distance between UE and BS} & \makecell[c]{$25-250 \rm m$} \\
    \hline
  \end{tabular}
  \label{parameters}
\end{table}

According to the 5G NR standard, the 132 resource blocks (RBs) occupying 1584 OFDM subcarriers near the center frequency are used for signal transmission, while the other OFDM subcarriers at two edges of the frequency band are used as null subcarriers. The pilot symbols are placed with equal interval on OFDM subcarriers. We consider two cases. In one case, we use $K=132$ pilot symbols, where each RB transmits a pilot symbol. In the other case, we use $K=88$ pilot symbols, where 3 RBs transmit 2 pilot symbols.

The performance of channel estimation can be evaluated by the normalized mean squared error (NMSE), which is defined as
\begin{equation}\label{NMSE}
  {\rm NMSE} = {\frac{\|\widehat{\bm{H}} - \bm{H}\|^2_F}{\|\bm{H}\|^2_F}}
\end{equation}
where $\widehat{\bm{H}}$ is an estimate of the genuine channel matrix $\bm{H}$.

\begin{figure}[!t]
  \centerline{\includegraphics[height=6.6cm]{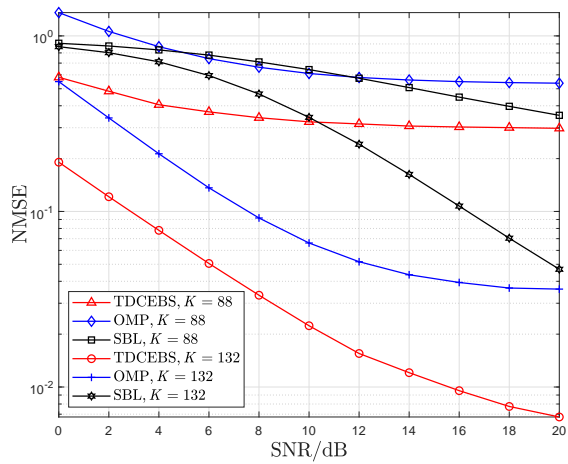}}
  \caption{Comparisons of channel estimation performance in terms of NMSE for different schemes.}
  \label{NMSEperformance}
\end{figure}

Fig.~\ref{NMSEperformance} shows the comparisons of the channel estimation performance in terms of NMSE for different schemes, including OMP, SBL~\cite{gred2019SBL} and the proposed TDCEBS scheme. It is seen that TDCEBS achieves the best performance, which is much better than both OMP and SBL. Either for $K=132$ or $K=88$, the NMSE of TDCEBS is significantly lower than that of OMP. When $\rm SNR=10dB$ and $K=132$, the NMSE of TDCEBS is reduced by about $\rm 4.7dB$ compared to that of OMP. When $\rm SNR=10dB$ and $K=88$, the NMSE of TDCEBS is about $\rm 2.7dB$ lower than that of OMP. Compared to TDCEBS, OMP makes sparse recovery independently for each column vector of $\bm H$, which does not exploit the block sparsity of $\bm H$. The SBL can only achieve the approximately-sparse channel estimation and there is channel power leakage for zero taps, which leads its performance to be worse than TDCEBS. For each scheme, using more pilots, e.g., from $K=88$ to $K=132$, can achieve better performance. In particular, such improvement is more apparent for TDCEBS and OMP than for SBL.

\begin{figure}[!t]
  \centerline{\includegraphics[height=6.6 cm]{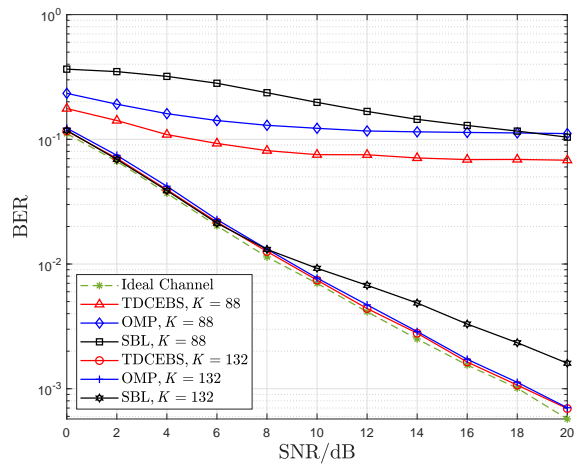}}
  \caption{Comparisons of BER performance for different schemes.}
  \label{BER}
\end{figure}

In addition to the NMSE performance, we also compare the bit error rate (BER) performance. Fig.\ref{BER} shows the BER performance comparisons among TDCEBS, OMP and SBL. To indicate the performance upper bound, we also include the BER performance when the ideal channel is known. It is seen that among the three schemes, TDCEBS performs the best. When $K=132$, both TDCEBS and OMP can well approach the performance upper bound. But when $K=88$, TDCEBS is substantially better than OMP. At $\rm SNR=10dB$, the improvement of TDCEBS over OMP is around $2.1{\rm dB}$, where the BER of TDCEBS might be reduced to zero if a proper channel coding scheme is applied. However, since OMP cannot achieve BER of 0.1 even at high SNR region, the channel coding scheme cannot be well applied for OMP in this context. Therefore, TDCEBS can function more effectively than OMP for channel estimation when the number of used pilot symbols is smaller.

\section{Conclusion}
In this letter, we have investigated time-domain channel estimation for wideband mmWave MIMO OFDM system. By transmitting frequency-domain pilot symbols as well as different beamforming vectors, we have observed that the time-domain mmWave MIMO channels exhibit channel delay sparsity and especially block sparsity among different spatial directions. We have proposed the TDCEBS scheme, which always aims at finding the best nonzero block achieving the largest projection of the residue at each iterations. In particular, we have evaluated the system performance using the QuaDRiGa which is recommended by 5G NR to generate wideband mmWave MIMO channels. Simulation results have verified the effectiveness of the proposed TDCEBS scheme and have shown that it can outperform the existing schemes. In the future work, we will continue to focus on the study of the sparse channel estimation exploiting joint delay and angle sparsity.

\bibliographystyle{IEEEtran}
\bibliography{IEEEabrv,IEEEexample}
\end{document}